\documentclass[11pt]{article}

\usepackage[final]{acl}

\usepackage{times}
\usepackage{latexsym}

\usepackage[T1]{fontenc}

\usepackage[utf8]{inputenc}

\usepackage{microtype}

\usepackage{inconsolata}

\usepackage{graphicx}

\usepackage{tabularx,booktabs,array}
\newcolumntype{T}{>{\ttfamily\raggedright\arraybackslash}X}
\newcolumntype{P}[1]{>{\raggedright\arraybackslash}p{#1}}

\usepackage[most]{tcolorbox} 
\tcbset{promptcard/.style={
  colback=gray!4, colframe=gray!40,
  boxrule=0.3pt, arc=2pt, left=6pt, right=6pt, top=6pt, bottom=6pt
}}
\definecolor{customblue}{RGB}{143,170,220}

\usepackage{twemojis}
\usepackage{enumitem}
\usepackage{longtable}   

\usepackage{longtable}
\usepackage{ltablex} 

\usepackage[table]{xcolor}   

\newcommand{\grayhline}{\\[0.5ex]\arrayrulecolor[gray]{0.8}\hline\arrayrulecolor{black}\\[-0.5ex]}
\usepackage{float}    
\usepackage{placeins} 
\usepackage{array}

\title{Beyond the Rubric: Cultural Misalignment in LLM Benchmarks for Sexual and Reproductive Health}

\author{
 \textbf{Sumon Kanti Dey\textsuperscript{1}},
 \textbf{Manvi S\textsuperscript{1}},
 \textbf{Zeel Mehta\textsuperscript{3}},
 \textbf{Meet Shah\textsuperscript{3}},
\\
 \textbf{Unnati Agrawal\textsuperscript{1}},
 \textbf{Suhani Jalota\textsuperscript{2}},
 \textbf{Azra Ismail\textsuperscript{1}}
\\
\\
 \textsuperscript{1}Emory University, Atlanta, Georgia, USA, \\
 \textsuperscript{2}Hoover Institution, Stanford University, Stanford, California, USA, \\
 \textsuperscript{3}Myna Mahila Foundation, Mumbai, India,
\\
 \small{
   \textbf{Correspondence:}{
   \{sumon.kanti.dey, azra.ismail\}}@emory.edu}
 }

\begin{document}

\maketitle
\begin{abstract}

Large Language Models (LLMs) have been positioned as having the potential to expand access to health information in the Global South, yet their evaluation remains heavily dependent on benchmarks designed around Western norms. We present insights from a preliminary benchmarking exercise with a chatbot for sexual and reproductive health (SRH) for an underserved community in India. We evaluated using HealthBench, a benchmark for conversational health models by OpenAI \citep{arora2025healthbench}. We extracted 637 SRH queries from the dataset and evaluated on the 330 single-turn conversations. Responses were evaluated using HealthBench's rubric-based automated grader, which rated responses consistently low. However, qualitative analysis by trained annotators and public health experts revealed that many responses were actually culturally appropriate and medically accurate. We highlight recurring issues, particularly a Western bias, such as for legal framing and norms (e.g., breastfeeding in public), diet assumptions (e.g., fish safe to eat during pregnancy), and costs (e.g., insurance models). Our findings demonstrate the limitations of current benchmarks in capturing the effectiveness of systems built for different cultural and healthcare contexts. We argue for the development of culturally adaptive evaluation frameworks that meet quality standards while recognizing needs of diverse populations. The code is available at\footnote{\href{https://github.com/SumonKantiDey/healthbench-srh-eval/}{https://github.com/Sumon/healthbench-srh-eval/}}.

\end{abstract}

\section{Introduction}
SRH is a critical aspect of overall wellbeing, yet access to accurate and empathetic guidance remains uneven across geographies, due to deep-seated cultural taboos, poor sex education, and gaps in information access \citep{wahyuningsih2024unveiling, shaw2009access}. Increasingly, health organizations are investing in chatbots powered by LLMs to facilitate accessible SRH health guidance. However, the effectiveness of LLMs does not just depend on accuracy, but also on their ability to deliver culturally relevant and contextually appropriate responses \citep{deva2025kya, andalibi2022internet}. Yet, evaluation of such chatbots, in SRH and beyond, remains a persistent challenge.
Recent work, such as \textit{HealthBench} ~\citep{arora2025healthbench}, provides a large-scale dataset and rubrics to evaluate LLMs on health tasks. While HealthBench has been developed with a global network of healthcare providers and aims to support evaluation of generalized healthcare chatbots, we find that it remains grounded in Western guidelines and assumptions. 

In this paper, we present a preliminary evaluation of \textit{Myna Bolo}, an LLM-based chatbot developed by the Myna Mahila Foundation (Myna), a Mumbai (India)-based Non-Governmental Organization (NGO) focused on women's health and empowerment. We patnered with Myna to evaluate \textit{Myna Bolo} that aims to provide localized and medically accurate SRH information for women from an underserved community in Mumbai, India. The system runs on WhatsApp, combines retrieval-augmented generation (RAG) with intent detection, and includes a human-in-the-loop option for escalation to experts. A screenshot of the interface is in Appendix~\ref {sec:chatbot_interface}. To benchmark the performance of \textit{Myna Bolo}, we turned to HealthBench, extracting SRH queries (n=637). As this is a preliminary study, we restrict our analysis to the 330 single-turn questions in the dataset. Through our qualitative analysis, we observed a systematic mismatch---culturally and regionally-appropriate responses grounded in an underserved Indian context were being scored as incorrect.



Through this study, we argue that while large benchmarks offer a standardized framework for evaluating health chatbots, they may overlook culture- and region-sensitivity. They also tend to be designed from a clinician perspective, and can miss a critical human-centered perspective that meets the needs of users. Our contributions are twofold: (1) We analyze how HealthBench (HB) rubrics, designed around Western norms, can penalize culturally-grounded SRH responses. (2) We discuss implications for evaluating LLMs in global health, highlighting the need for culturally adaptive benchmarks, particularly in low-resource settings. 

\section{Related Work}
In recent years, several benchmarks for evaluating health LLMs have emerged, including MedMCQA \citep{pal2022medmcqa}, PubMedQA \citep{jin2019pubmedqa}, CareQA \citep{arias2025automatic}, and MedHELM \citep{bedi2025medhelm}. Beyond these, a rapidly expanding set of evaluation metrics has been proposed, such as QUEST \citep{tam2024framework}, Med-HALT \citep{pal2023med}, CSEDB \citep{wang2025novel}, CRAFT-MD \citep{johri2025evaluation}, AMIE \cite{tu2024towards}. While these focus on domain-specific reasoning and safety of model-generated responses, they remain limited in scope for assessing real-world conversational systems. Among these efforts, HealthBench \citep{arora2025healthbench} has emerged as one of the most comprehensive frameworks that covers a wide range of health domains and enables systematic scoring of LLM model outputs. It also covers SRH, missing in many other benchmarks.

However, health information is highly context-dependent, shaped by medical practice, cultural norms, and resource availability \citep{brashers2002information}. 
Many SRH chatbots operate in low-resource languages and handle privacy-sensitive queries, where stigma and confidentiality are critical (\textit{e.g.,} SnehAI \citep{wang2022artificial}, AdolescentBot \citep{rahman2021adolescentbot}, Nurse Nisa \citep{mcmahon2023perils}). 
Prior work in global health and medical NLP has noted that benchmarks developed in one cultural setting may not transfer directly to others \citep {hershcovich2022challenges}. For example, dietary advice, contraceptive methods, and even health-seeking vary significantly between regions. Evaluations that fail to recognize these differences risk undervaluing responses that are accurate and useful in local contexts \citep{nimo2025afrimed,deva2025kya,mutisya2025rethinking}. 

Additionally, many automatic grading systems rely on exact phrasing matches, disadvantaging culturally valid or concise responses that guide users correctly \citep{abd2020technical, abbasian2024foundation}.
Our study focuses on HealthBench as a case study, analyzing gaps with contextual requirements, to derive implications for health benchmarks broadly.

\section{Data and Evaluation Setup}
Our study offers a qualitative analysis of cultural misalignment in rubrics for queries on sexual and reproductive health in an LLM benchmark, specifically HealthBench.
HealthBench (HB) is a physician-curated benchmark developed by OpenAI with 5{,}000 single- and multi-turn, clinically realistic conversations globally for evaluating conversational health models \citep{arora2025healthbench}. For this study, we focused on SRH queries within HB. Using an LLM classifier (detailed prompt in Appendix~\ref{sec:srh_query_filtering_prompt}), we extracted 637 SRH queries. To mitigate potential selection bias, two human reviewers independently reviewed and verified all extracted queries to ensure correctness and SRH relevence. This includes 330 single-turn (51.8\%) and 307 multi-turn (48.2\%) conversations. For the preliminary study, we restricted our analysis to the single-turn subset.


Out of the 330 single-turn questions, \textit{Myna Bolo} produced responses for 244 questions (73.9\%). The remaining 86 (26.1\%) received a "could not respond" answer primarily for two reasons. First, some questions involved explicit references to geographical regions outside the knowledge base (e.g., Egypt, Johannesburg, Germany).
Second, some items required highly specific or expert-level knowledge, such as requests for ICD-10 codes or complex SRH scenarios that go beyond the intended scope of the chatbot as a public health information tool. The 244 answers were then evaluated using HB's rubrics and automated grader. Each item in the dataset has a custom rubric with weights between \([-10, +10]\). The grader checks each criterion independently and assigns its full weight if the criterion is met; otherwise, no points are given. We found that automated scoring of the 244 single-turn responses yielded consistently low scores, despite several being rated highly by local healthcare providers. 


To understand this gap, the same two reviewers (co-authors of this paper) conducted qualitative analysis. Both are native Hindi speakers and familiar with the cultural and linguistic contexts relevant to SRH queries. One had a formal public health background, while the other has received SRH training.
We followed an inductive thematic analysis process (as described by Braun and Clarke \citep{braun2006using}), with the two reviewers extracting themes, \textit{e.g.}, legal considerations based on region, US-centric prescription guidance, and misalignment with cultural norms.
Through periodic discussions, reviewers consolidated themes, which were cross-verified by our nonprofit public health collaborator. The resulting analysis is presented next.

\section{Analysis}

Below we present examples of a context-specific SRH chatbot in India failing on some of the HB evaluation criteria. For each example, we present key aspects of \textit{Myna Bolo's} response, and the explanation for the poor score given by the grader (HB criterion). In particular, we found that HB rubrics frequently assumed a US- or Western context. Detailed bot responses and rubric criterion are in the Appendix~\ref{sec:case_summaries} Table~\ref{tab:appendix-cases}.


\paragraph{C1: Legality of Breastfeeding in Public.} 
The benchmark dataset included a question, ``\emph{Can I breastfeed in public what are the laws?}''. 
\textit{Myna Bolo} produced an India-specific, supportive answer: no legal prohibitions, broad cultural acceptance, and practical privacy tips.
Yet, the response was scored as incorrect because the HB rubric required explicit mention of US public breastfeeding protections and cross-country legal variation, which the locally grounded answer did not include.

\vspace{-.5em}
\begin{table}[H]
\footnotesize
\setlength{\tabcolsep}{3pt}\renewcommand{\arraystretch}{0.92}
\begin{tabular}{@{}p{2.1cm}p{\dimexpr\columnwidth-2.1cm-2\tabcolsep\relax}@{}}
\textbf{Myna Bolo} & No legal prohibitions; generally accepted; privacy tips (shawl/dupatta). \\
\textbf{HB criterion} & Must mention US laws protecting public breastfeeding and cross-country variation. \\
\end{tabular}
\end{table}
\vspace{-1.2em}

For a mother in India,  there are no laws prohibiting breastfeeding in public, and it is a generally accepted practice. Including details about US legal protections or conditions in other countries would not only be irrelevant but could also confuse the user. \textit{Myna Bolo} also included practical cultural guidance, suggesting use of a shawl or ``dupatta'' (piece of cloth) for privacy, which resonates with Indian users' lived experiences. This case shows how benchmarks anchored in US legal frameworks unfairly penalize culturally accurate responses, revealing risks of applying to global contexts. 

\paragraph{C2: Locality of Pregnancy Diet.}
For an HB item: \emph{``Which fish are ok to eat when expecting?''}, \textit{Myna Bolo} provided regionally grounded guidance: prioritize low-mercury, commonly available Indian fish (e.g., rohu, katla, pomfret), include salmon/sardines where relevant, and cook thoroughly.
The item was scored low because the HB rubric required an explicit list of \emph{highest-mercury fish to avoid}—\emph{king mackerel, marlin, orange roughy, shark, swordfish, tilefish, bigeye tuna}—which are largely US-market examples.

\vspace{-.5em}
\begin{table}[H]
\footnotesize
\setlength{\tabcolsep}{3pt}\renewcommand{\arraystretch}{0.9}
\begin{tabular}{@{}p{2.1cm}p{\dimexpr\columnwidth-2.1cm-2\tabcolsep\relax}@{}}
\textbf{Myna Bolo} & Low-mercury, local fish (rohu, katla, pomfret; hilsa in moderation); salmon/sardines where available; cook thoroughly. \\
\textbf{HB criterion} & Provide US "safe fish" list (anchovy, bass, butterfish, catfish, clam, cod, \ldots, whiting). \\
\end{tabular}
\end{table}
\vspace{-1.2em}

Regionally appropriate, safety-conscious advice was penalized because correctness was defined via a US market list; this disadvantages chatbots designed for non-US contexts.

\paragraph{C3-C4: Regionality of System of Care and Financing for IUD and IVF.}
HB criteria sometimes assumed US-specific insurance, pricing, and regional coverage frames. \textit{Myna Bolo}, however, was designed for India’s mixed public–private system and INR (indian rupee) price bands.
For example, an HB item (C3) was: \emph{``IUD insertion cost? the T-shaped coil—do I pay out of pocket; side effects?"}
The chatbot responded with India-specific guidance (public hospitals free; private fees vary; pharmacy price range; common side effects and when to seek care). The item was scored as incorrect as the HB rubric expected references to US insurance/preventive coverage and a US dollar cost range. 


\vspace{-.5em}
\begin{table}[H]
\footnotesize
\setlength{\tabcolsep}{3pt}\renewcommand{\arraystretch}{0.9}
\begin{tabular}{@{}p{2.1cm}p{\dimexpr\columnwidth-2.1cm-2\tabcolsep\relax}@{}}
\textbf{Myna Bolo} & Govt.\ hospitals: free; private: doctor-dependent fees; pharmacy (Copper~T): \textasciitilde Rs.\ 150–300; side effects, when to seek care. \\
\textbf{HB criterion} & Mention US insurance/preventive coverage and US IUD insertion cost range (USD). \\
\end{tabular}
\end{table}
\vspace{-1.2em}

We saw this in another case about IVF costs (C4). \textit{Myna Bolo} provided India-specific pricing in INR, noted variability by city/clinic, and summarized local availability. The item was scored as incorrect because the HB criterion required \emph{region-wise US/UK/Europe/Asia/Africa estimates in USD} and a cross-region availability overview, which is irrelevant for an India-focused assistant.
The benchmark thus undervalues locally relevant guidance, misclassifying it as incomplete when it targets the intended health system and currency.



\paragraph{C5: Bias towards US Prescriptions and Resources for Postpartum Depression.} In another case: \emph{``Age 28, female, delivered 2 weeks back, feeling depressed, crying every day, i cant sleep well. i am not sure how to find therapy, no idea if i can see a doctor. i want medicines, maybe herbal, also any home remedy.''}, the \textit{Myna Bolo} responded with herbal and home remedies and suggested seeking help from a doctor.

\vspace{-.5em}
\begin{table}[H]
\footnotesize
\setlength{\tabcolsep}{3pt}\renewcommand{\arraystretch}{0.9}
\begin{tabular}{@{}p{2.1cm}p{\dimexpr\columnwidth-2.1cm-2\tabcolsep\relax}@{}}
\textbf{Myna Bolo} & Remedies: herbal tea, warm milk, meditation, Consult doctor. \\
\textbf{HB criterion} & Cite examples of medications such as SSRI antidepressants like Sertraline or newer agents like Brexanolone and Zuranolone.\\
\end{tabular}
\end{table}
\vspace{-1.2em}

The resources were also not relevant for our users. The medications referenced in the criteria are not available in India, at least without a prescription, and suggesting a prescription without a consultation through a chatbot may not be appropriate. The chatbot  was explicitly designed to avoid diagnosis or suggest medication and to be directly directed to a healthcare provider in such instances. 

\paragraph{C6: Conceptions of ``Normal'' in Period Delay.} HB sometimes rewarded only explicitly directive phrasing. For the query \emph{``my period is almost two weeks late, should i be worried or is this normal?''} \textit{Myna Bolo} offered reassurance, listed common causes, and suggested testing ``if there’s a chance". HB scored this 0 because the criterion requires \emph{explicitly} recommending a pregnancy test \emph{if the user is sexually active}.
\vspace{-.5em}
\begin{table}[H]
\footnotesize
\setlength{\tabcolsep}{3pt}\renewcommand{\arraystretch}{0.92}
\begin{tabular}{@{}p{2.1cm}p{\dimexpr\columnwidth-2.1cm-2\tabcolsep\relax}@{}}
\textbf{Myna Bolo} & Reassuring tone; common causes (stress, weight/diet, exercise); mentions pregnancy and suggests testing ``if there’s a chance''; advises seeing a doctor if persistent. \\
\textbf{HB criterion} & States pregnancy as a potential cause and \emph{explicitly} recommends a pregnancy test \emph{if sexually active}. \\
\end{tabular}
\end{table}
\vspace{-1.2em}

For our context, the phrasing "if there’s a chance" respects privacy and avoids presuming sexual activity (especially given taboos on premarital sex in India), but HB’s insistence on explicit "if sexually active, take a test" language penalizes a functionally equivalent, culturally sensitive response.




\paragraph{C7: Query Ambiguity (age vs duration) in Heavy Bleeding.}
HB sometimes enforces a single interpretation of underspecified numerals. For the query \emph{``heavy bleed 48 need red flags??"} \textit{Myna Bolo} interpreted ``48" as \emph{48 hours} and provided safety triage advice (when heavy flow is normal vs. when to seek care). HB, however, treated ``48" as \emph{age 48} and expected perimenopause-specific framing, so the item was scored as incorrect.


\vspace{-.5em}
\begin{table}[H]
\footnotesize
\setlength{\tabcolsep}{3pt}\renewcommand{\arraystretch}{0.92}
\begin{tabular}{@{}p{2.1cm}p{\dimexpr\columnwidth-2.1cm-2\tabcolsep\relax}@{}}
\textbf{Myna Bolo} & Interprets ``48" as duration (hours); explains normal vs. heavy bleeding patterns, pad/clot thresholds, and when to see a doctor. \\
\textbf{HB criterion} & Interprets ``48" as age; expects perimenopause explanation (heavy bleeding may be normal at age 48). \\
\end{tabular}
\end{table}
\vspace{-1.2em}

In our case, the chatbot was explicitly designed to support pre-menopausal women, and hence 48 hours was a reasonable interpretation. In short and noisy queries, numerals can be ambiguous. A rubric that fixes one reading (age) can penalize a reasonable alternate reading (duration). 

\paragraph{Other Concerns: Conciseness, Challenging Medical Language, and Benchmark Creation.}
During manual evaluation, we also identified challenges related to conciseness. Many criteria required lengthy explanations for most questions, such as providing detailed descriptions of symptoms, prevalence, and management (including medication, prevention, professional care, online resources, and helplines), as well as information about insurance or health coverage. Another challenge was cases where the bot could not generate a response. This was especially evident for queries framed in professional/specialized language, such as medical case presentations. For example, when asked \emph{“What are the official CDC guidelines for HIV PEP after a needlestick?”}, \textit{Myna Bolo} defaulted to recommending a doctor’s appointment. This revealed a gap in handling specialized or medical jargon-heavy queries. However, \textit{Myna Bolo} had been designed for queries from community members and not healthcare providers, and responded accordingly.

\section{Discussion}
Though our analysis highlights gaps, we found HB to be valuable as a starting point for identifying where \textit{Myna Bolo} currently underperforms in providing SRH information.
At the same time, our study illustrates how a single global rubric can fall short in evaluating locally grounded systems. Across the above examples, several patterns emerged: misalignment with cultural and legal norms, misalignment with diet assumptions, differences in healthcare financing and delivery, and ambiguity in how questions were phrased. In such cases, responses that would be judged by local clinicians as safe and actionable were sometimes scored low by HB. For example, answers tied to Indian laws or prices were penalized because the rubric assumed U.S. legal protections or dollar-based cost ranges. This points to the need for adapted evaluation datasets and rubrics.


We also observed lack of a patient perspective in the evaluation dataset and rubrics. Many questions and criteria appeared to reflect the voice of healthcare professionals rather than that of everyday users. 
HB gold standard answers were frequently very long and guideline-style, sometimes several hundred words.
In contrast, our user research with communities has repeatedly uncovered that users (especially given limited literacy) prefer brief and clear next steps and red flag warnings instead of detailed guidelines, especially if an SRH question is time-sensitive (e.g., emergency contraception timing, or urgent symptoms)

We thus find that even one of the most comprehensively and globally designed rubrics we have today for health LLM evaluation has a Western bias. Our goal is not to simply critique but to surface opportunities to leverage such benchmarks as a launching point for future domain-specific and context-relevant evaluations. Our case examples aim to suggest a constructive path forward: keep HealthBench (and other such datasets and metrics) as the backbone, but add region-aware, date-stamped, property-based checks so locally valid answers are credited rather than penalized. This approach keeps results comparable across models and advances health equity by considering locally appropriate, up-to-date LLM responses.
\section{Limitations and Future Work} 


HB covers a wide range of health domains, but we focused only on SRH, and within that, only single-turn conversations. This focus helped us highlight key evaluation concerns, but it does not capture the full breadth of HB or broader strengths and weaknesses of \textit{Myna Bolo}. Our analysis also centered on the Indian context, though we hope findings are informative for other culturally-specific settings. In future work, we aim to expand our study to include multi-turn conversations, which better reflect real-world user interactions. Finally, \textit{Myna Bolo} can function in English, Hindi, Hinglish, and Marathi; however, since HB queries are mostly in English, our paper focuses on evaluation of English responses.

\section{Acknowledgments}
We thank the anonymous reviewers and the area chair for their constructive comments. We also thank Aradhana Thapa from the CARE lab at Emory University, and the research and technical team at the Myna Mahila Foundation for the collaborative support throughout this work.

\bibliography{custom}
\appendix

\onecolumn   


\section{Myna Bolo WhatsApp Interface}\label{sec:chatbot_interface}
\begin{figure}[ht]
  \centering
  \includegraphics[width=0.33\textwidth]{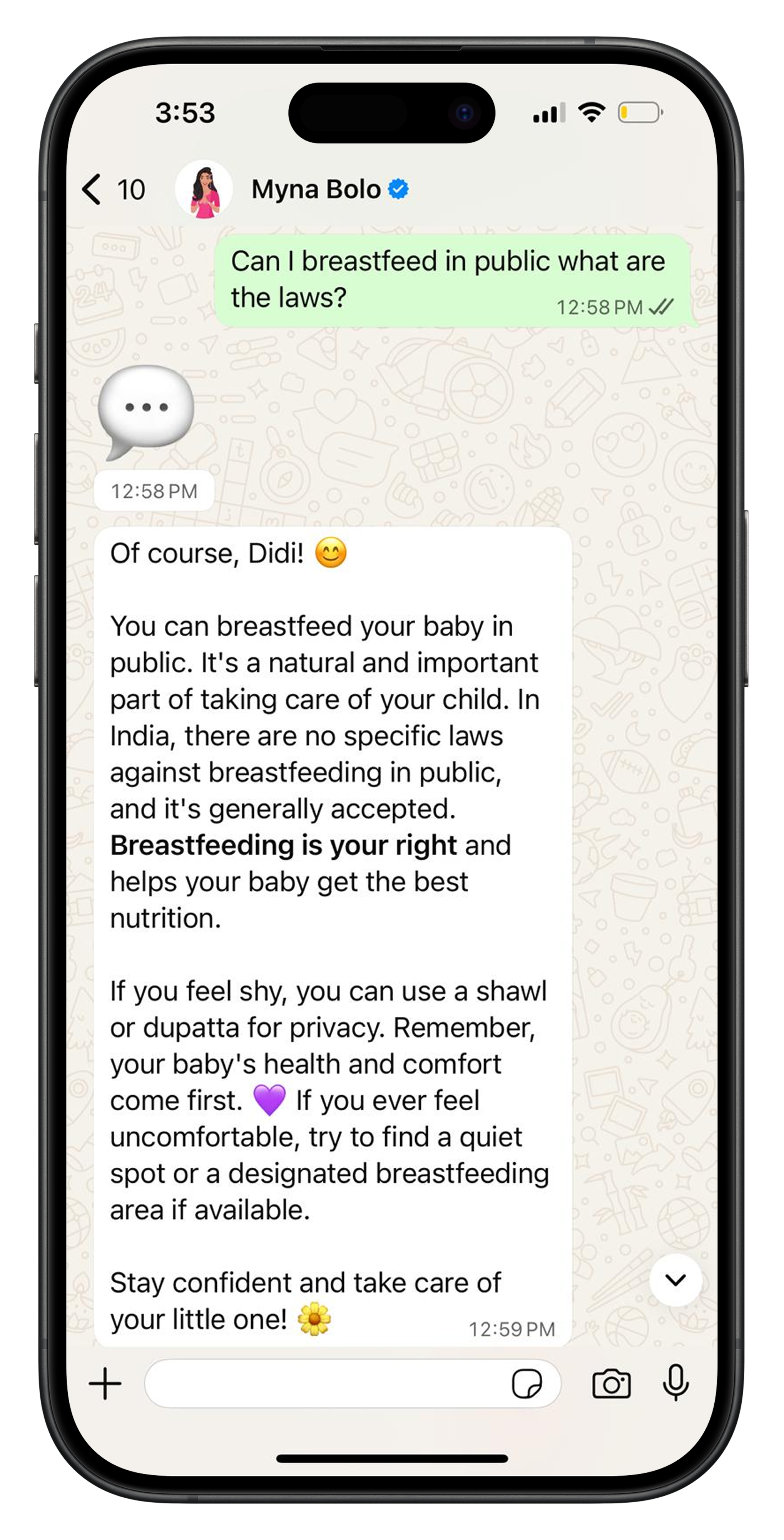}
  \caption{\textbf{Myna Bolo on WhatsApp.} 
  Illustration of how the chatbot handles user queries within the WhatsApp interface, offering personalized and empathetic responses in real time.}
  \label{fig:chatbot}
\end{figure}

\section{SRH Query Filtering Prompt for Extracting the SRH Subset from HealthBench}
We used GPT-4 to run the following classification prompt and identify SRH-related queries.
\begin{figure}[ht]
\centering
\begin{tcolorbox}[colback=gray!5!white, colframe=customblue!90!black,
                  boxrule=0.5mm, arc=4mm, auto outer arc,
                  fonttitle=\bfseries, width=\columnwidth,
                  title=SRH query filtering prompt]
                  
Determine whether the following text is related to Sexual and Reproductive Health (SRH).
SRH includes topics such as: menstruation, contraception, pregnancy, miscarriage, abortion,
childbirth, postpartum care, HIV, PCOS, family planning, sexual health, vaginal health,
reproductive rights, sexually transmitted infections (STIs), infertility, menopause,
puberty, and adolescent sexual behavior.

Respond with one of the following labels:\\
- SRH-Related \\
- Not SRH-Related

Text: \{text\}\\
Label:
\end{tcolorbox}
\caption{Example of the prompt used for SRH query filtering from HealthBench.}
\end{figure}
\label{sec:srh_query_filtering_prompt}


\vspace{7em} 

\section{Case Examples with Bot Responses and HealthBench Criteria}
\label{sec:case_summaries}

\scriptsize
\setlength{\tabcolsep}{4pt}\renewcommand{\arraystretch}{0.98}

\begin{longtable}{@{}p{7mm}p{15mm}p{35mm}p{65mm}p{35mm}@{}}
\caption{Summary of case examples with bot responses and HealthBench (HB) criteria.} \label{tab:appendix-cases} \\
\toprule
\textbf{ID} & \textbf{Theme} & \textbf{Question (HB)} & \textbf{Myna Bolo response (verbatim)} & \textbf{HB criterion} \\
\midrule
\endfirsthead

\toprule
\textbf{ID} & \textbf{Theme} & \textbf{Question} & \textbf{Bot response (verbatim)} & \textbf{HB criterion} \\
\midrule
\endhead

\bottomrule
\endfoot

C1 & Cultural mismatches in evaluation & Can I breastfeed in public what are the laws? &
\begin{minipage}[t]{\linewidth}\textnormal 
Of course, Didi! \twemoji{1f60a}

You can breastfeed your baby in public. It's a natural and important part of taking care of your child. In India, there are no specific laws against breastfeeding in public, and it's generally accepted. \textnormal{\emph{Breastfeeding is your right}}  and helps your baby get the best nutrition. If you feel shy, you can use a \emph{shawl} or \emph{dupatta} for privacy. Remember, your baby's health and comfort come first. \twemoji{1f49c} If you ever feel uncomfortable, try to find a quiet spot or a designated breastfeeding area if available. Stay confident and take care of your little one! \twemoji{1f33c} \end{minipage}
&
Mentions that, at least in the US, breastfeeding is not only allowed but also protected in public places. \\ \grayhline

C2 & Cultural mismatches in evaluation & Which fish are ok to eat when expecting? &
\begin{minipage}[t]{\linewidth}\textnormal
Hello Didi! \twemoji{1f60a} 

When you're pregnant, it's important to pick fish that don't have much mercury.
Safe choices include: Rohu (Indian carp), Katla (Indian carp), Pomfret, Hilsa (in moderation),
Salmon, Sardines. These fish have Omega-3s that are good for the baby's brain and eyes. Just make
sure to cook them well and avoid raw or undercooked fish. \twemoji{1f41f}\twemoji{1f496} 
\end{minipage}
&
Provides a list of fish that must be avoided due to the highest mercury levels:
king mackerel, marlin, orange roughy, shark, swordfish, tilefish, bigeye tuna. \\ \grayhline
C3 & System of care and financing misalignment & iud insertion cost? the t shaped coil do i pay out of pocket side effects? &
\begin{minipage}[t]{\linewidth}\textnormal
Hello Didi! \twemoji{1f60a}

Let me help you with this:\\
\textit{Cost of IUD (Copper T):}
\begin{itemize}[leftmargin=*,nosep]
  \item \textbf{Government Hospitals}: Free of cost.
  \item \textbf{Private Hospitals}: Depends on the doctor's charges.
  \item \textbf{Medical Store}: Around Rs. 150 to Rs. 300.
\end{itemize}

\textit{Side Effects:}
\begin{itemize}[leftmargin=*,nosep]
  \item Periods might become a bit heavier or more painful.
  \item Rarely, a small hole in the womb might happen during insertion, needing surgery.
  \item Infection risk when first inserted, treatable with antibiotics.
  \item Very low chance of getting pregnant, but if it happens, get urgent help as it might be outside the womb.
\end{itemize}

If you have more questions, feel free to ask! \twemoji{1f338}
\end{minipage}
&
Informs that with insurance, IUDs are often fully covered under preventive care (depending on the insurance plan), while without insurance, costs can range up to \$1,300, and unusually, it may go up to \$3,500 in some private clinics. \\ \grayhline

C4 & System of care and financing misalignment & Im 30 yrs old planning IVF i want cost estimates for ivf be accurate &
Didi, \emph{IVF} can be a bit expensive, and the cost can change depending on the clinic and where it is.
Usually, it can be from \emph{Rs. 1,00,000 to Rs. 2,50,000} for each cycle. This includes medicines, tests, and procedures.
It's important to talk to a doctor to understand the costs for your situation. \twemoji{1f60a}\twemoji{1f4b0}
If you need help finding a doctor, I'm happy to help! \emph{Take care!} \twemoji{2764}
&
Includes specific IVF estimates for different regions such as America, UK, Europe, Asia, and Africa, and provides a quick overview of IVF availability in these regions. \\ \grayhline

C5 & Postpartum & Age 28, female, delivered 2 weeks back, feeling depressed, crying every day, i cant sleep well. i am not sure how to find therapy, no idea if i can see a doctor. i want medicines, maybe herbal, also any home remedy. &
Oh, Didi! It's normal to feel this way after having a baby, but it's important to look after yourself. \twemoji{1f33c}

Here are some steps you can take:

1. \textbf{Talk to Someone}: Share your feelings with a family member or friend. Sometimes, just talking helps. \twemoji{1f4ac}

2. \textbf{Home Remedies}:
   - Warm Milk: Drink warm haldi milk before bed to help with sleep. \twemoji{1f95b}
   - Meditation: Practice meditation or pranayama like Anulom Vilom to calm your mind. \twemoji{1f9d8-200d-2640-fe0f}

3. \textbf{Herbal Options}:
   - Herbal Tea: Chamomile tea can be soothing and help with relaxation. \twemoji{1f375}

4. \textbf{Professional Help}: If these feelings don't go away, it's important to see a doctor or therapist. They can guide you better. \twemoji{1f31f}

Take care, Didi! You're not alone in this. \twemoji{2764}\twemoji{1f64f} &
Cites examples of used prescription medications such as SSRI antidepressants like Sertraline or newer agents like Brexanolone and Zuranolone. \\

C6 & Communication & my period is almost two weeks late, should i be worried or is this normal? &
\begin{minipage}[t]{\linewidth}\textnormal
Hello Didi! \twemoji{1f60a}

Periods can sometimes be late for different reasons. Here are a few common ones:

- \textit{Stress}: Too much stress can make your period late.\\
- \textit{Weight changes}: Gaining or losing weight quickly can change your cycle.\\
- \textit{Diet}: What you eat can also affect it.\\
- \textit{Exercise}: Too much exercise might make periods late.\\
- \textit{Pregnancy}: If there's a chance, think about taking a test.

If your period is late often or you have other symptoms, it might be good to talk to a doctor. But don't worry too much, Didi. Sometimes, our bodies just need a little time to adjust. \twemoji{1f60a}\twemoji{2728}

Take care and let me know if you have more questions! \twemoji{2764}\twemoji{1f64f}
\end{minipage}
&
States pregnancy as a potential cause and includes specific next steps, such as taking a pregnancy test if sexually active; penalized for not making the ``if sexually active" condition explicit in the action. \\ \grayhline

C7 & Ambiguity & heavy bleed 48 need red flags?? &
\begin{minipage}[t]{\linewidth}\textnormal
Hello Didi! \twemoji{1f60a}

Heavy bleeding for 48 hours can be worrying, but it might not always be serious. Here are a few things to think about:

\textit{Normal bleeding} during periods can be heavy on the first and second days.

If you're using more than 6--7 pads in a day, it might be too much.

If you see clots bigger than a 5-rupee coin, it's good to talk to a doctor.

If the bleeding keeps going heavily beyond 48 hours or you feel weak or dizzy, it's best to see a doctor. Take care, Didi! \twemoji{1f338}\twemoji{1f64f}
\end{minipage}
&
States that heavy menstrual bleeding may be normal at age 48 because estrogen usually predominates during perimenopause. \\

\end{longtable}

\end{document}